\documentclass[%
 reprint,
superscriptaddress,
 nofootinbib,
 nobibnotes,
 amsmath,amssymb,
 aps,
 pra,
 floatfix,
]{revtex4-1}

\usepackage{graphicx}% Include figure files
\usepackage{dcolumn}% Align table columns on decimal point
\usepackage{bm}% bold math
\usepackage{hyperref}% add hypertext capabilities
\usepackage[ngerman,english]{babel}
\usepackage[utf8]{inputenc}
\usepackage[T1]{fontenc}
\usepackage{xspace}
\newcommand{\dd}{\ensuremath{\mathrm{d}}\xspace}
\newcommand{\ls}{\ensuremath{\ell^\star}\xspace}
\newcommand{\kls}{\ensuremath{k\ell^\star}\xspace}

\usepackage{xcolor}
\begin{document}

%\preprint{APS/123-QED}

\title{%
Resonant transport and near-field effects in photonic glasses
}% Force line breaks with \\

\author{Geoffroy J. \surname{Aubry}}%
\thanks{These authors contributed equally to this work.}
%\email{geoffroy.aubry@uni-konstanz.de}
\affiliation{%
 Fachbereich Physik, Universität Konstanz, 78457 Konstanz, Germany
}%
\author{Lukas \surname{Schertel}}
\thanks{These authors contributed equally to this work.}
\affiliation{%
 Fachbereich Physik, Universität Konstanz, 78457 Konstanz, Germany
}%
\affiliation{
 Physik-Institut, Universität Zürich, 8057 Zürich, Switzerland
}%
\author{Mengdi \surname{Chen}}%
\affiliation{%
 Fachbereich Chemie, Universität Konstanz, 78457 Konstanz, Germany
}%
\author{Henrik \surname{Weyer}}%
\affiliation{%
 Fachbereich Physik, Universität Konstanz, 78457 Konstanz, Germany
}%

\author{Christof M. \surname{Aegerter}}
\affiliation{
 Physik-Institut, Universität Zürich, 8057 Zürich, Switzerland
}%
\author{Sebastian \surname{Polarz}}%
\affiliation{%
 Fachbereich Chemie, Universität Konstanz, 78457 Konstanz, Germany
}%
\author{Helmut \surname{Cölfen}}%
\affiliation{%
 Fachbereich Chemie, Universität Konstanz, 78457 Konstanz, Germany
}%
\author{Georg \surname{Maret}}
\affiliation{%
 Fachbereich Physik, Universität Konstanz, 78457 Konstanz, Germany
}%

\date{\today}% It is always \today, today,

\begin{abstract}
A fundamental quantity in multiple scattering is the transport mean free path the inverse of which describes the scattering strength of a sample.
In this paper, we emphasize the importance of an appropriate description of the effective refractive index $n_{\mathrm{eff}}$ in multiple light scattering to accurately describe the light transport in dense photonic glasses.
Using $n_{\mathrm{eff}}$ as calculated by the energy-density coherent potential approximation we are able to predict the transport mean free path of monodisperse photonic glasses.
This model without any fit parameter is in qualitative agreement with numerical simulations and in fair quantitative agreement with spectrally resolved coherent backscattering measurements on new specially synthesized polystyrene photonic glasses. These materials exhibit resonant light scattering perturbed by strong near-field coupling, all captured within the model.
Our model might be used to maximize the scattering strength of high index photonic glasses, which are a key in the search for Anderson localization of light in three dimensions.
\end{abstract}

\maketitle

Transport phenomena are omnipresent in nature, governing many processes in chemistry, biology, physics, and engineering.
Systems as diverse as electrons~\cite{Abrahams1979} and ultrasound~\cite{Bayer1993} in condensed matter, mechanical waves in the earth~\cite{Larose2004}, cold atoms in an optical trap~\cite{Jendrzejewski2012a}, and light in disordered photonic materials~\cite{Wolf1985,*Albada1985} share the same physical principle~\cite{MesoPhys, Lagendijk2009}.
Optical experiments are especially appealing because of the absence of photon-photon interaction (unlike in electronic systems) and the existence of relatively high index scattering media such as photonic crystals and glasses.
Moreover, optical transport experiments have reached an unprecedented accuracy thanks to the great technological development of sources (e.g., lasers), detectors (e.g., CCDs), and time resolution.
All these progresses allow the realization of table-top experiments which highlight the richness of transport phenomena.

\begin{figure}
\includegraphics[width=\columnwidth]{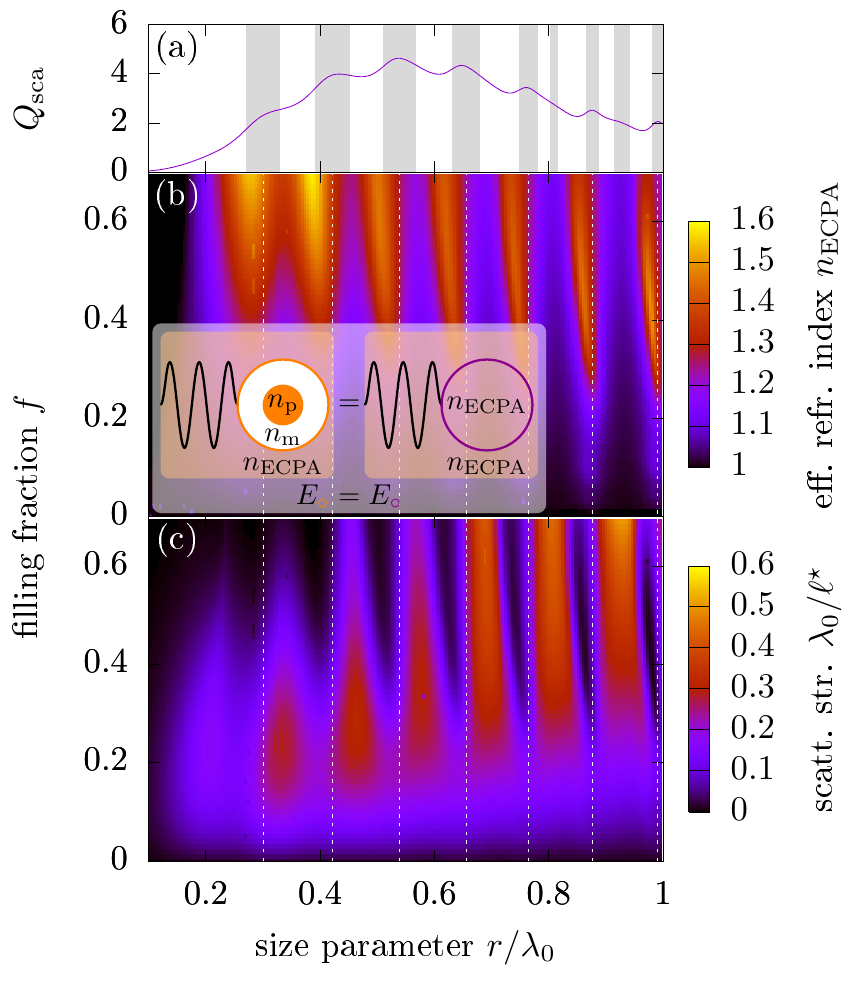}
\caption{(a) Scattering efficiency $Q_\mathrm{sca}$ of an isolated Mie scatterer having a refractive index of $n_\mathrm{PS}=1.6$ in air. The shaded regions highlight the region with negative curvature and the dashed white lines are a guide to the eye. -- (b) Color map of $n_{\mathrm{ECPA}}$ calculated for a refractive index of the particle $n_{\mathrm{PS}}$ surrounded by air. Inset: principle of the ECPA (see text) --
(c) Color map of the scattering strength $\lambda_0/\ls$ calculated with Eq.~(\ref{eq:lstar}) using $n_{\mathrm{ECPA}}$.
In both cases, the quantities are shown as a function of both the size parameter $r/\lambda_0$ and the filling fraction $f$.}
\label{fig:PS-maps}
\end{figure}

Wave transport in a diluted disordered suspension of scatterers can be described by the sole far-field properties of the single scatterers.
On increasing concentration, however, interference effects due to scatterer-scatterer position correlation need to be taken into account~\cite{Fraden1990}.
In this description, the scattering cross section is still the single scatterer one, which, in general, is calculated in the far field.
In optics, this approach is expected to fail as soon as the photon scattering mean free path $\ell_\mathrm{s}$ becomes smaller than a few wavelengths of the light.
In this case, the distance between two scattering events is so short that (1) each scattered photon does not reach the far field limit before being re-scattered and (2) the (differential) scattering cross section of each and every scatterer is affected by multiply scattered photons returning to it.
A first attempt to describe these near-field effects was recently proposed by \textcite{RezvaniNaraghi2015}, but takes into account only the first point.
In this paper, we propose a different light transport model that takes into account both effects by considering that each scatterer in a densely packed random assembly is embedded in a properly estimated effective medium~\footnote{The importance of an appropriate description of the effective refractive index to describe light transport in randomly packed colloids was already noted by \textcite{Reufer2007}.}.
We show that the use of the energy-density coherent potential approximation (ECPA, \cite{Busch1995,*Busch1996}) for the calculation of the effective refractive index provides a proper description of light transport in densely packed multiple scattering samples.
The predictions of this model are tested experimentally on samples having resonant transport properties, namely, monodisperse photonic glasses, which were first introduced and studied by García and coworkers~\cite{Garcia2007,Sapienza2007,Garcia2008}.
These materials are made of randomly assembled monodisperse Mie scatterers (which are isolated dielectric spheres of the same radius $r$).
A common idea is that single scattering Mie resonances (see Fig.~\ref{fig:PS-maps}(a) or Ref.~\onlinecite{Bohren1998}) have a signature in the transport mean free path $\ls$ of randomly assembled Mie scatterers.
These resonances were already observed, but up to now, all attempts to describe them more than qualitatively failed~\cite{Sapienza2007,Reufer2007,Montesdeoca2016}.
We argue here that a proper description of resonant near-field effects is the key to a quantitative description of them.

In this paper, we present a fit parameter free model that is able to predict for the first time the correct positions ($r/\lambda_0$) of the resonance peaks of the scattering strength ($\lambda_0/\ls$) and describes their order of magnitude.
We tested this model against \textit{ab initio} numerical simulations, earlier experimental data obtained from transmission experiments~\cite{Garcia2008}, and new experimental results from backscattering experiments on specially synthesized polystyrene colloidal glasses~\cite{Chen2017}.

Our description of the light transport in photonic glasses starts with the derivation of the transport mean free path~\cite{Wolf1988},
\begin{equation}
\ls= \frac{\ell_\mathrm{s}}{1-\langle \cos \theta \rangle} \overset{\text{sphere}}{=}  \frac{1}{1-\langle \cos \theta \rangle  }\frac{4 \pi r^3}{3f \sigma_{\mathrm{s}}}. \label{eq:lstar}
\end{equation}
The second equality holds in the case of an assembly of monodisperse Mie scatterers.
$\ell_\mathrm{s}$ is the scattering mean free path (the distance between two scattering events), $f$ is the filling fraction of the scatterers, and $\theta$ is the scattering angle.
Both the scattering cross section $\sigma_\mathrm{s}$ and the anisotropy factor $\langle \cos \theta \rangle$ of each scattering event in the multiple scattering regime can be expressed in terms of the form factor $F(\theta)$ (scattering properties of the single sphere) and the structure factor $S(\theta)$ (collective scattering of the sample)~\cite{Fraden1990, Hulst1957}
\begin{equation}
\sigma_\mathrm{s} = \frac{\pi}{k^2}\int_0^\pi F(\theta) S(\theta) \sin \theta \,\dd\theta,
\label{eq:scattCS}
\end{equation}
\begin{equation}
\langle \cos \theta \rangle= \dfrac{\int_0^\pi \cos \theta F(\theta) S(\theta) \sin \theta \,\dd\theta}{ \int_0^\pi F(\theta) S(\theta) \sin \theta \,\dd\theta }.
\label{eq:asyPara}
\end{equation}
The intensity form factor $F(\theta)$ of a sphere is calculated by the Mie-theory~\cite{Bohren1998}.
Since the samples we focus on are assemblies of randomly packed spheres, we use the hard sphere Percus-Yevick structure factor~\cite{Percus1958} $S(q)$ with $q=2k\sin\theta/2$ for $S(\theta)$ ($k=2\pi/\lambda$ with $\lambda$ the wavelength of the light in the surrounding medium).
Using this description, \textcite{Fraden1990} were able to describe the effect of  short-range correlations of the scatterers positions on multiple scattering of light in polystyrene spherical colloids ($r=230$\,nm) suspended in water with filling fractions up to 45\% (refractive indices $n_\mathrm{PS}=1.60$ in $n_\mathrm{H_2 O}=1.33$).

However, when the average distance between nearest colloids is of the order of the light wavelength, near-field effects start to play a role in the transport properties~\cite{RezvaniNaraghi2015}.
These near-field effects are not caught in $F(\theta)$ when calculated with the bulk refractive index of the surrounding medium.
Moreover, $S(q)$ depends on the surrounding refractive index via $\lambda$ as well.
As we show below, the presence of other scatterers in the direct vicinity can be taken into account by defining an effective refractive index $n_{\mathrm{eff}}$ for the surrounding medium.
$n_{\mathrm{eff}}$, which can first be roughly estimated as the volume average of the local refractive index, lowers the refractive index difference between each scatterer and the medium, and therefore lowers its scattering strength.
This effect was not considered in Ref.~\onlinecite{Fraden1990}, but does not alter much that analysis because of the relatively low refractive index contrast between polystyrene and water.
We show here that a proper estimation of $n_{\mathrm{eff}}$ is necessary when dealing with higher refractive index contrasts (like polystyrene in air, $n_0=1$)~\footnote{Let us just mention here that we are not aware of any direct measurement of this quantity.}.

To estimate the effective refractive index of our photonic glasses, we use the ECPA by \textcite{Busch1995}
\footnote{A comparison between the ECPA and other ways of defining $n_\mathrm{eff}$~\cite{Albada1991,Soukoulis1994,Soukoulis1989,Economou1989} is given in Appendix~\ref{sec:effectiveIndex}.}.
Since the positions of the scatterers are random (with the constraint that the spheres can not overlap), each one will be coated, on average, by a shell having the refractive index of the embedding matrix. This \emph{core-shell} particle itself is embedded in a medium having the effective refractive index (see inset of Fig.~\ref{fig:PS-maps}(b)).
By definition, the energy density inside an effective medium should be uniform: we find $n_\mathrm{ECPA}$ using an iterative process such that the energy density inside the core-shell particle matches the energy density inside the same volume, but having the uniform refractive index $n_\mathrm{ECPA}$.
The result of this iterative process is plotted in Fig.~\ref{fig:PS-maps}(b) as a function of the size parameter and of the scatterer density for polystyrene colloids embedded in air.
Clear resonances are seen as a function of the size parameter.
We can furthermore see an increase of the refractive index with increasing filling fraction.
We note, finally, that the largest $n_{\mathrm{ECPA}}$ corresponds approximately to the index of the scattering particles ($n_{\mathrm{PS}}=1.6$).

\begin{figure}
\includegraphics[width=\columnwidth]{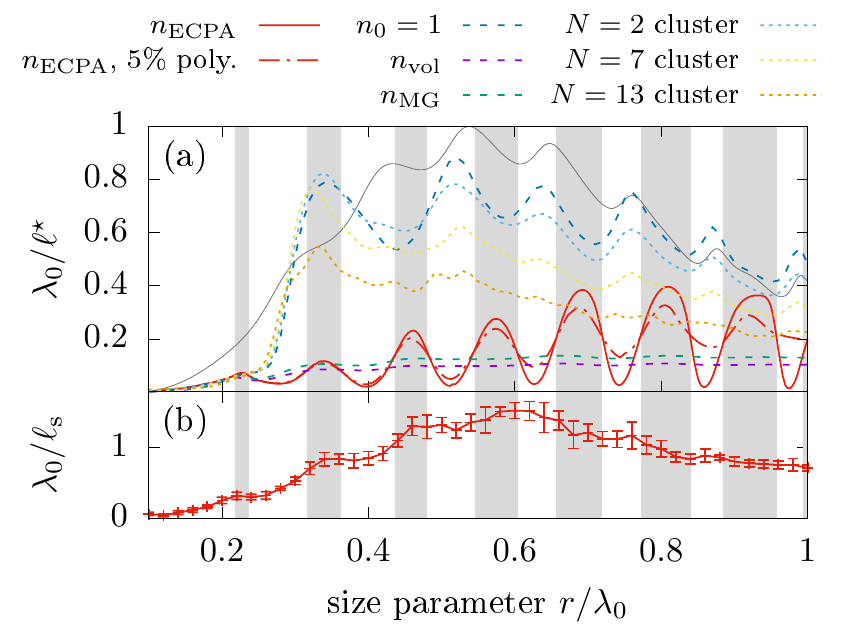}
\caption{(a) Scattering strength $\lambda_0/\ls$ calculated with different models for a filling fraction $f=0.5$ (red curves: using $n_\mathrm{ECPA}$ (solid: monodisperse, dot dashed: 5\% polydispersity; the shaded regions highlight negative curvature of the solid red curve); dashed curves: simple model for $n_\mathrm{eff}$ (see text); dotted curves: calculated with the differential cross section of the $N$ cluster). The single sphere scattering efficiency (also plotted in Fig.~\ref{fig:PS-maps}(a) is shown for comparison in gray (arbitrary units). --
(b) Averaged $\lambda_0/\ell_\mathrm{s}$ calculated with the MSTM code over five realizations of numerical photonic glasses, each one having $f=0.5$.}
\label{fig:PS-overlstar-simulations}
\end{figure}

Replacing the wave vector $k$ by $k_{\mathrm{eff}}=2\pi n_{\mathrm{eff}}/\lambda_0$ in the calculation of $\sigma_\mathrm{s}$ and $\langle \cos \theta \rangle$, we can calculate the scattering strength $\lambda_0/\ls$ using Eq.~(\ref{eq:lstar}) as a function of both $r/\lambda_0$ and $f$ (see Fig.~\ref{fig:PS-maps}(c)).
Again resonances can be seen as a function of the size parameter.
Note that the peaks of $\lambda_0/\ls$ are placed at dips of $n_{\mathrm{ECPA}}$.
This can be understood in terms of partial index matching between the scatterers and the effective surrounding medium.
Fig.~\ref{fig:PS-overlstar-simulations}(a) shows the scattering strength for a given filling fraction ($f=0.5$, red solid curve) and compares it to the one expected taking a given polydispersity of the spheres into account (dash-dotted red curve).
As expected, the resonances are somewhat smeared out.

\paragraph*{Comparison with other models:} To emphasize our claim that a proper choice of $n_{\mathrm{eff}}$ is necessary for the calculation of the transport properties of dense high index photonic glasses, we now show the difference between the scattering strength calculated using $n_{\mathrm{ECPA}}$ (red solid curve in Fig.~\ref{fig:PS-overlstar-simulations}) and the one calculated using more simple models ($n_{\mathrm{vol}}=f\cdot n_{\mathrm{PS}}+ (1-f) \cdot n_0$ (violet) or $n_\mathrm{MG}$ for the Maxwell Garnett effective refractive index~\cite{Garnett1904} (green)).
The simple models lead to no resonant behavior in $n$ but the order of magnitude of the scattering strength agrees more with the $n_\mathrm{ECPA}$ predictions than the curve calculated without using any effective refractive index (blue dashed curve).

An other way to take the presence of other scatterers in the direct vicinity of each very scatterer into account for its scattering properties is to calculate directly $F(\theta)$ for a scatterer surrounded by other scatterers (as suggested by \textcite{Garcia2008}).
We therefore calculated numerically using the Multiple Sphere T Matrix (MSTM) code~\cite{Mackowski2011} the average over all possible orientations of the differential scattering cross section $F_N(\theta,d_\mathrm{NN}(f))$ of clusters of $N=2$, 7, or 13 particles made of one sphere having $N-1$ neighboring spheres placed at the average nearest-neighbor distance in the glass $d_\mathrm{NN}$ (which is a function of $f$~\cite{Torquato1995}).
We then replaced $F(\theta)$ by $F_N(\theta,d_\mathrm{NN}(f))/N$ in Eqs.~(\ref{eq:scattCS}) and (\ref{eq:asyPara}) \emph{without} using an effective refractive index.
By doing this, we take into account near-field effects in the scattering properties in a similar way as in Ref.~\onlinecite{RezvaniNaraghi2015}.
As shown by the dotted curves in Fig.~\ref{fig:PS-overlstar-simulations}(a), the  scattering strength decreases with the number of nearest neighbors indicating stronger near-field effects, but is still different than the one predicted using $n_\mathrm{ECPA}$.

\paragraph*{Numerical test:} To test the model using $n_\mathrm{ECPA}$, we simulate the transmission trough thin slabs (6.25 times the radius of the spheres) of monodisperse spheres using the MSTM code~\cite{Mackowski2011}.
The glasses have a filling fraction of 50\%, and were numerically synthesized by using a force-biased algorithm~\cite{Baranau2014}.
In our simulations, we calculated the transmission of the non-scattered field and extracted the \emph{scattering} mean free path from this quantity~\footnote{See Appendix~\ref{sec:numSim}.}.
As expected, the so-extracted $\ell_\mathrm{s}$ are smaller than the $\ls$ predicted by our model.
Moreover, one can recognize the same resonant behavior between $\ell_\mathrm{s}$ and $\ls$, at least for $r/\lambda_0\lesssim0.7$.
For larger ones, the resonances seems to be smeared out (in this range, the overall transmissions fall in the $10^{-3}$ range, making it difficult to extract the coherent beam without any scattered contribution).

\begin{figure}
\includegraphics[width=\columnwidth]{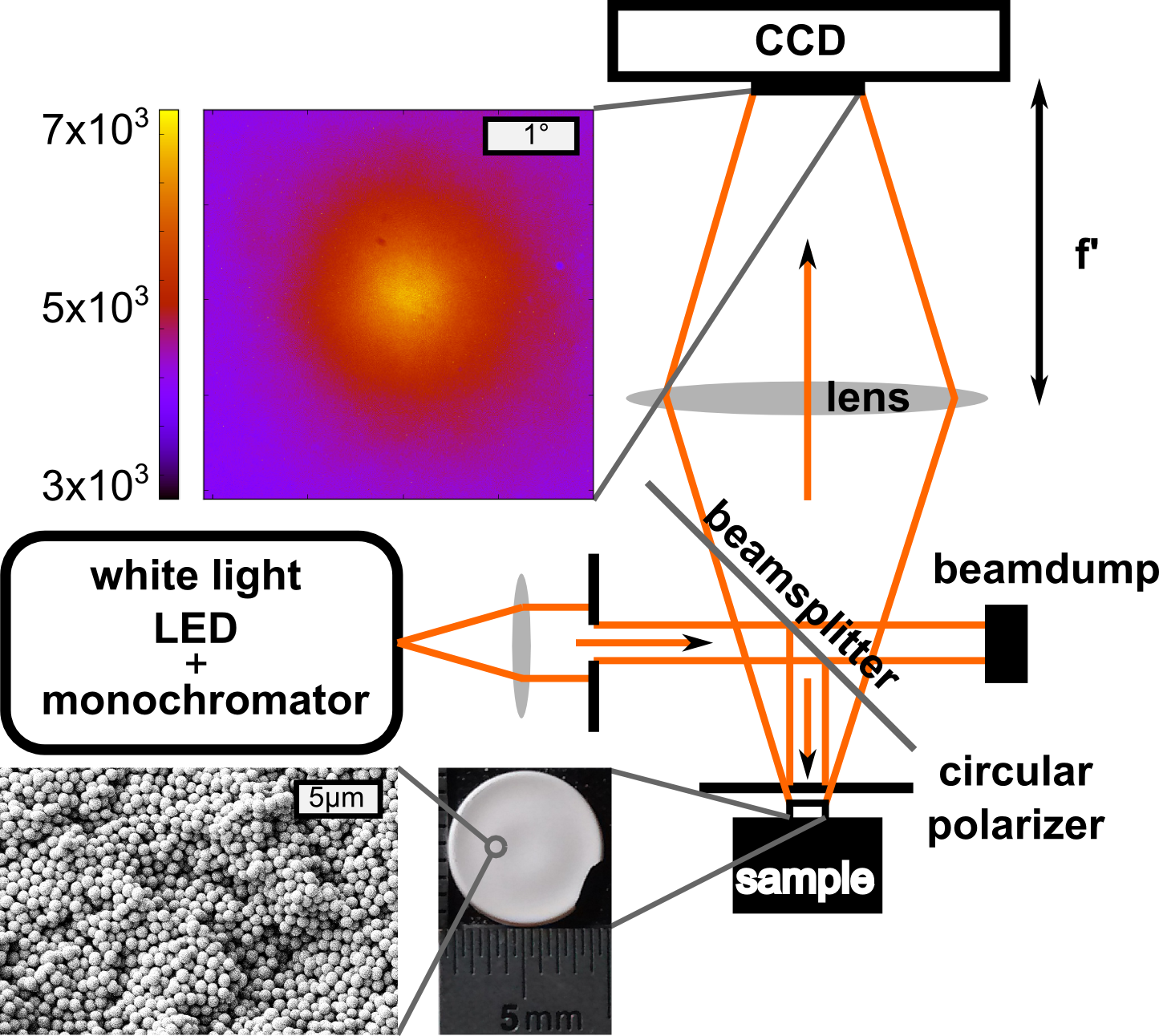}
\caption{Scheme of the CBC setup used to measure $\ls$ of photonic glasses: Illumination can be tuned from 450\,nm to 700\,nm via a white LED coupled in a monochromator.
The output illuminates the sample via a 50:50 beamsplitter and a circular polarizer is used to filter single scattered light. The reflected light is imaged in the focal plane of a $200$\,mm lens on a CCD camera.
A typical image of the cone of a polystyrene photonic glass is shown on the left of the CCD. An image of a macroscopic polystyrene sample is shown in the lower left with a SEM image of the same sample to its left.}
\label{fig:Setup}
\end{figure}

\paragraph*{Experimental test:}
To test the model experimentally, we measure the transport mean free path $\ls$ of photonic glasses for different size parameters $r/\lambda_0$ by analyzing the shape of their coherent backscattering cone (CBC)~\cite{Wolf1985,Albada1985}. The width of the CBC is inversely proportional to $\kls$.
In this method a parallel light beam illuminates the multiple scattering sample via a beamsplitter (see Fig.~\ref{fig:Setup}), and the backscattered light is imaged in the Fourier space on a CCD camera (Apogee Alta U4000) placed in the focal plane of a convex lens ($f'=200\,\mathrm{mm}$).
A circular polarizer in front of the sample filters single backscattered light~\footnote{In single scattering events the helicity is flipped and filtered while multiple scattering randomizes the polarization.}.
This setup is able to measure angles up to $3.0^{\circ}$.
To probe $\ls$ for different $r/\lambda_0$ ratios, we used a white LED~\footnote{Lumileds LUXEON CoB 109 5000\,K} coupled in a monochromator~\footnote{Acton SP-2150i, Princeton Instruments} as light source to tune $\lambda_0$ between 450 and 700\,nm with a wavelength width of $\Delta \lambda_0 \approx 5\,\mathrm{nm}$. Since this is an incoherent source, there is no need to average over speckles to measure the CBC~\footnote{This light source has an slightly divergent beam which needs to be accounted for in the analysis, see Appendix~\ref{sec:divergence}.}.
Our samples are free standing photonic glasses having a thickness $L\approx 1\,\mathrm{mm}$ and a diameter $d\approx 1\,\mathrm{cm}$. The samples were prepared from monodisperse polystyrene colloids in solution via ultracentrifugation and with addition of a controlled amount of salt~\cite{Chen2017} to avoid crystallization and prevent optical shortcuts. A small amount of polyacrylamide holds the particles together to allow freestanding photonic glasses.
Therefore, from the measurement of the mass and of the volume of the samples, we can only estimate an upper limit for the filling fractions of the particles in the glasses. Their filling fraction might be slightly lower than 0.55.
With our light source and by varying the particles size from $r=125$ to $r=335\,\mathrm{nm}$~\footnote{The radii and their errors are obtained from the analysis of SEM images (as the one shown in Fig.~\ref{fig:Setup} for the sample with the largest particles).
All samples have polydispersities lower than $5\%$.} we are able to cover a wide range of $r/\lambda_0$ from 0.17 to 0.70.

\begin{figure}
\includegraphics[width=\columnwidth]{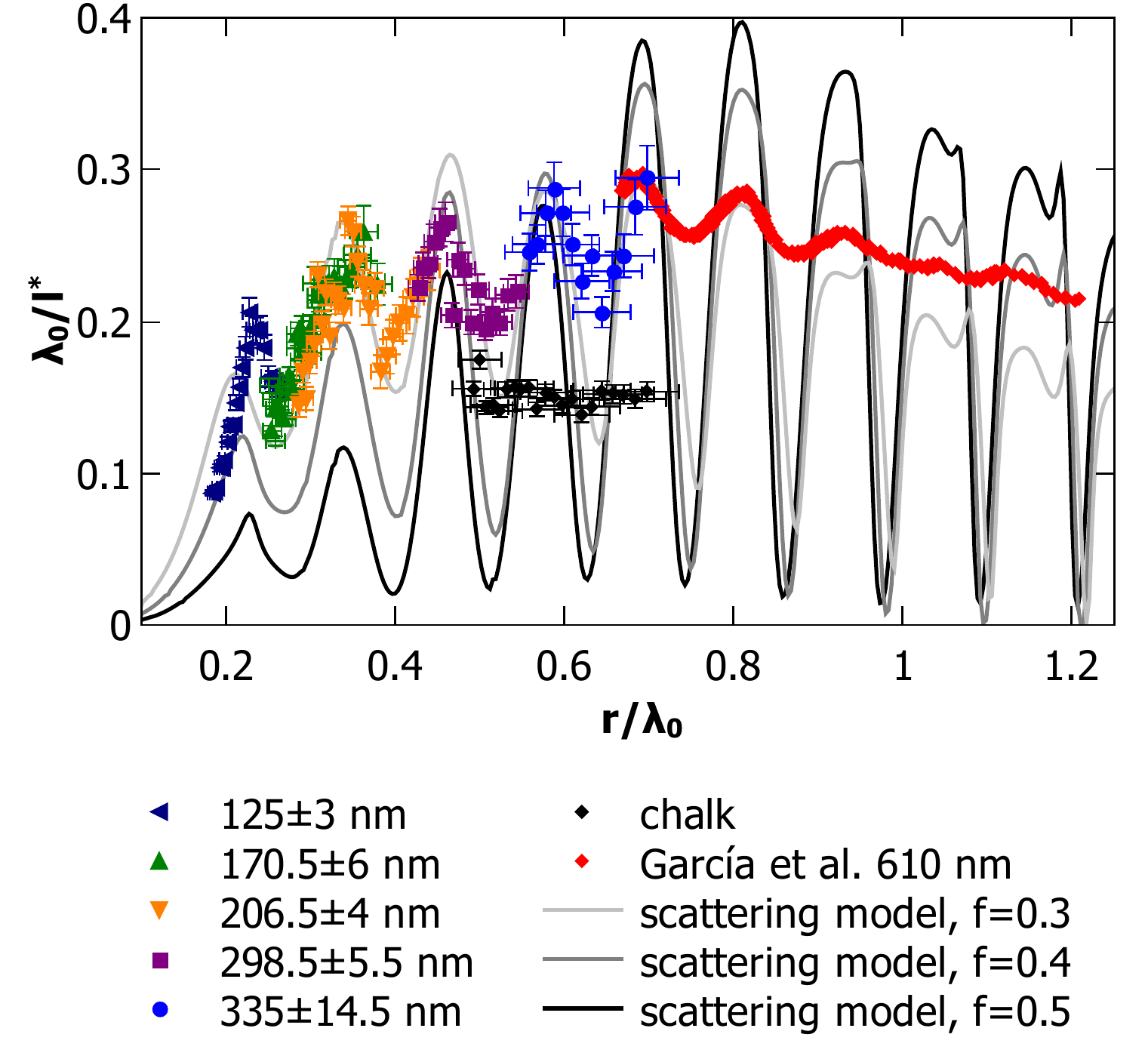}
\caption{
Measured scattering strength of five monodisperse polystyrene photonic glasses from $r=125$ to $335\,\mathrm{nm}$ compared to a chalk sample (irregular particle shape, for this sample we arbitrarily set $r=335$\,nm).
The errors in $\ls$ are estimated from the fits to be $\pm 0.1\,\mu \mathrm{m}$.
The data are compared with the predictions of the scattering model (Eq.~(\ref{eq:lstar}) using $n_\mathrm{ECPA}$) for different filling fractions (solid lines) and with the transmission data extracted in \textcite{Garcia2008} ($r=610$\,nm).}
\label{fig:Measurement}
\end{figure}

The image in Fig.~\ref{fig:Setup} shows the CBC of a polystyrene photonic glass.
For each measurement we averaged about five images (exposure time $t_{\mathrm{exp}}=3\,\mathrm{s}$) before performing a radial average.
The obtained data were finally fitted with the standard CBC formula to obtain $\ls$~\cite{MesoPhys}.
The diffusion constant $D$ and the absorption time $\tau_{\mathrm{a}}$ were measured separately in time of flight experiments to extract the absorption length $L_{\mathrm{a}}=\sqrt{D \tau_{\mathrm{a}}}$ which is used in the CBC fit.
$L_{\mathrm{a}}$ was typically on the order of the sample thickness $L\gg \ls$, meaning that absorption was very low.
A wavelength scan for five different photonic glasses is shown in Fig.~\ref{fig:Measurement} and reveals the first direct observation of strong resonances of $\lambda_0/\ls$ in the visible.
For comparison we measured a randomly shaped and highly polydisperse sample of chalk powder (black diamonds) and, as expected, no resonant behavior is observed in this case.
The experimental data show that the positions of the resonances are very well predicted when the transport properties are calculated using the ECPA effective refractive index (see the comparison between the scattering strengths predicted by different models in Fig.~\ref{fig:PS-overlstar-simulations}).
The amplitudes of the resonances are slightly smeared out compared to the model.
This can partially be understood because of the residual polydispersity of the particles and of the wavelength width of the source (see Fig.~\ref{fig:PS-overlstar-simulations}).
To further test our model we plotted the data of Fig.~19 of~\textcite{Garcia2008} (red diamonds).
They measured $\ls$ via diffuse transmission measurements in the infrared with polystyrene particles of $r=610\,\mathrm{nm}$.
Our model recovers the position of their resonances very well.
The lower amplitude of the measured resonances compared with the model could be explained by the polydispersity of their sample, but no such information is given in the paper.
All these six different data sets collapse on the same curve and cover the first nine Mie resonances versus $r/\lambda_0$, and this without any fit parameter.

In this paper we highlight the importance of an appropriate model for $n_{\mathrm{eff}}$ in the description of the transport properties of dense photonic materials where near-field coupling influences scattering behavior.
The shell introduced in the definition of $n_\mathrm{ECPA}$ is the key to take near-field effects into account.
It has a thickness which is related to the average particle distance in the glass, and couples electromagnetically each scatterer with the surrounding medium.
Unlike all other models used so far, the ECPA model predicts very well the resonant behavior in photonic glasses without any fit parameter.

The resonant behavior in polymeric photonic glasses leads to strong scattering at certain $r/\lambda_0$ values.
Increasing the refractive index contrast by using higher index materials such as TiO$_2$ spheres in air will lead to much stronger resonances. While the preparation of freestanding photonic glasses made of such high index colloids with the right $r/\lambda_0$ values may still be a technical issue, it would possibly give access to a new scattering regime where signatures of three-dimensional light localization are expected.

\begin{acknowledgments}
The authors would like to thank Ralf Tweer for providing his ECPA-code and Herbert Kaiser for useful input on the numerical part of this paper.
We further thank the Center for Applied Photonics (CAP) at the Universität Konstanz and the Schweizerischer Nationalfonds (SNF) for financial support.
The numerical simulations were done on the Scientific Compute Cluster at the Universität Konstanz.
GJA acknowledges support from the Zukunftskolleg (Universität Konstanz) for an Independent Research Start-up Grant.
\end{acknowledgments}

\appendix
\section{Effective refractive index}\label{sec:effectiveIndex}

Beside the need of an effective index theory that takes into account the resonant behavior of Mie scatterers (like the subject of this paper), different effective medium theories were developed in the quest for an appropriate description of transport properties in the high concentration regime in multiple scattering samples.

The energy transport velocity $v_\mathrm{E}$ of a wave propagating in a random medium is lowered with respect to the speed of light in vacuum, $v_{\mathrm{E}}=\frac{c}{n_{\mathrm{eff}}}$, $n_{\mathrm{eff}}$ being the effective refractive index of the complex material.
For very small concentrations of scattering particles $n_{\mathrm{eff}}$ tends towards the refractive index of the matrix medium $n_{\mathrm{m}}$. If the sample is completely filled by the particle medium $n_{\mathrm{eff}}$ converges to $n_{\mathrm{p}}$.
The easiest way to define an effective refractive index which captures this behavior is
\begin{equation}
n_{\mathrm{eff}}=f n_{\mathrm{p}} + (1-f) n_{\mathrm{m}}.
\label{eq:Simple}
\end{equation}
This is only valid for extreme situations of very low or very high filling fractions $f$.

Another way to calculate the effective refractive index is the Maxwell Garnett mixing formula~\cite{Garnett1904}
\begin{align}
n_\mathrm{MG}=n_{m}\sqrt{1+\frac{3fK}{1-fK}}
\label{eq:MG}
\end{align}
with $K=\frac{n_{\mathrm{p}}^2-n_{\mathrm{m}}^2}{n_{\mathrm{p}}^2+2n_{\mathrm{m}}^2}$.
The latter model is more physical than Eq.~(\ref{eq:Simple}) because it assumes that the polarizabilities are additive, not the refractive indexes.
Nevertheless, even if the derivation of Eq.~(\ref{eq:MG}) assumes the scatterers to be spherical, it completely neglects resonant scattering (the fact that energy can be stored in Mie scatterers for certain radius over wavelength ratios).

By using a Bethe-Salpeter equation, \textcite{Albada1991} were able to describe the effect of these Mie resonances in the regime of low $f$, and their theory was consistent with experiments at $f=0.36$. 
This showed that experimentally obtained low values of the diffusion constant $D=\frac{v_\mathrm{E} \ls}{3} $ were related to low values of $v_\mathrm{E}$ due to resonant Mie scattering and a corresponding energy storage process, and not to low values of $\ls$ which would signify localization~\cite{Busch1995,*Busch1996}.
For higher filling fractions (like in the case studied in this paper) experimental data are much better predicted by advanced versions of the so called coherent potential approximation (CPA)~\cite{Soukoulis1994}.

\subsection{Coherent Potential Approximation}

In the ``genuine'' CPA~\cite{Soukoulis1989,*Economou1989} a medium of two lossless materials consisting of spheres with radius $r$, refractive index $n_{\mathrm{p}}$, and a volume fraction $f$ in a host material with refractive index $n_{\mathrm{m}}$ is considered.
Each point in the medium will be either in a region with $n_{\mathrm{p}}$ with a probability $f$, or in a region with $n_{\mathrm{m}}$ with a probability $1-f$.
The medium is modeled by spheres of radius $r$ having a refractive index given by the aforementioned probabilities.
The effective refractive index is then found such that the averaged forward scattering amplitude vanishes,
\begin{align}
f F_\mathrm{p}(0) + (1-f) F_\mathrm{m}(0) = 0,
\label{eq:CPA}
\end{align}
with $F_\mathrm{p}$ (resp. $F_\mathrm{m}$) being the differential scattering cross-section of a sphere of refractive index $n_\mathrm{p}$ (resp. $n_\mathrm{m}$) embedded in a medium having the effective refractive index.

This approach neglects topological and geometrical differences between the scattering spheres and the host material, e.g., for high $f$ the host spheres are not only less probable but also have to have smaller radii. The real random system would be much better estimated by a mixture of coated sphere (where the scatterer with $n_{\mathrm{p}}$ is coated by a spherical region of host material $n_{\mathrm{m}}$) and of spheres (of refractive index $n_{\mathrm{m}}$).
Equation~(\ref{eq:CPA}) (this time with $F_\mathrm{p}$ the differential scattering cross-section of the coated sphere) is then solved in the same way as in the classical CPA to obtain the effective medium refractive index $n_{\mathrm{eff}}$ self-consistently.
In this approach the coating thickness varies with $f$. Due to the condition that the spheres should not overlap, the distribution of spacings between neighboring spheres has a peak at $r_\mathrm{c}>r$ with $r_\mathrm{c}=\frac{r}{f^{1/3}}$. This advanced version of the CPA has the advantage of taking into account short range order and thus fits to experimental data quite well in the high $f$ regime~\cite{Soukoulis1994}.
Nevertheless, the effective refractive index calculated by the coated CPA is very close to the one calculated with the Maxwell-Garnett theory (see Fig.~4 of Ref.~\onlinecite{Soukoulis1994}), whereas Fig.~\ref{fig:PS-overlstar-simulations}(a) shows that the Maxwell-Garnett theory does not fit our data.

\subsection{Energy-Density Coherent Potential Approximation}
In the classical CPA the energy density is homogeneous by construction. This can be violated in the coated CPA approach because of the coated sphere as the basic scattering unit. For low $f$ (large coatings) this leads to unphysical behavior near the single sphere Mie resonances, e.g., refractive indices smaller than 1 such that $v_\mathrm{E}>c$. Therefore a new CPA approach was developed by \textcite{Busch1995}, the so-called energy-density coherent potential approximation.
Here the termination criterion for the iterative determination of the effective refractive index $n_{\mathrm{eff}}$ is that a homogeneous energy density $\rho_\mathrm{E}$ on scales larger than the basic scattering unit is reached.
This is schematically shown in the inset of Fig.~\ref{fig:PS-maps}(b).

The criterion of a constant energy density in the case of a plane wave hitting a coated sphere embedded in the effective medium versus the case where the same volume is filled by the effective medium only is quantitatively expressed in the self-consistent equation
\begin{equation}
\int_0^{r_\mathrm{c}} \dd^3R \rho^{(1)}_{\mathrm{E}}(\vec{R})= \int_0^{r_\mathrm{c}} \dd^3R \rho^{(2)}_{\mathrm{E}}(\vec{R}) \ ,
\end{equation}
where $\rho^{(1)}_{\mathrm{E}}(\vec{R})$ and $\rho^{(2)}_{\mathrm{E}}(\vec{R})$ are the energy densities in the coated sphere and in the same volume filled with a medium having a refractive index $n_\mathrm{eff}$ respectively.
The energy density of a electromagnetic wave can be expressed as:
\begin{equation}
\rho_{\mathrm{E}}(\vec{R})=\frac{1}{2}\left[\varepsilon(\vec{R}) \vert \vec{E}(\vec{R}) \vert ^2 + \mu \vert \vec{H}(\vec{R}) \vert ^2 \right] \ .
\end{equation}
where $\vec{E}$ and $\vec{H}$ are the electric and magnetic fields, $\varepsilon$ is the dielectric constant, and $\mu$ is the magnetic permeability (the latter is assumed to be the same in both materials). With these equations the effective refractive index can be determined for all frequencies guaranteeing a homogeneous energy density on scales larger than the scattering unit.

\textcite{Busch1995} claim that their model, which takes into account multiple scattering effects in a mean-field sense, can be used for scalar (acoustic), vector (electromagnetic), and tensor (elastic) waves and is valid for all densities of scatterers.
They also test it on earlier experimental data.
More interesting for the present paper, this model captures the effects of resonant near-field effects in the multiple scattering regime.

\section{Numerical simulations}\label{sec:numSim}

\begin{figure}
\begin{center}
\includegraphics[width=\linewidth]{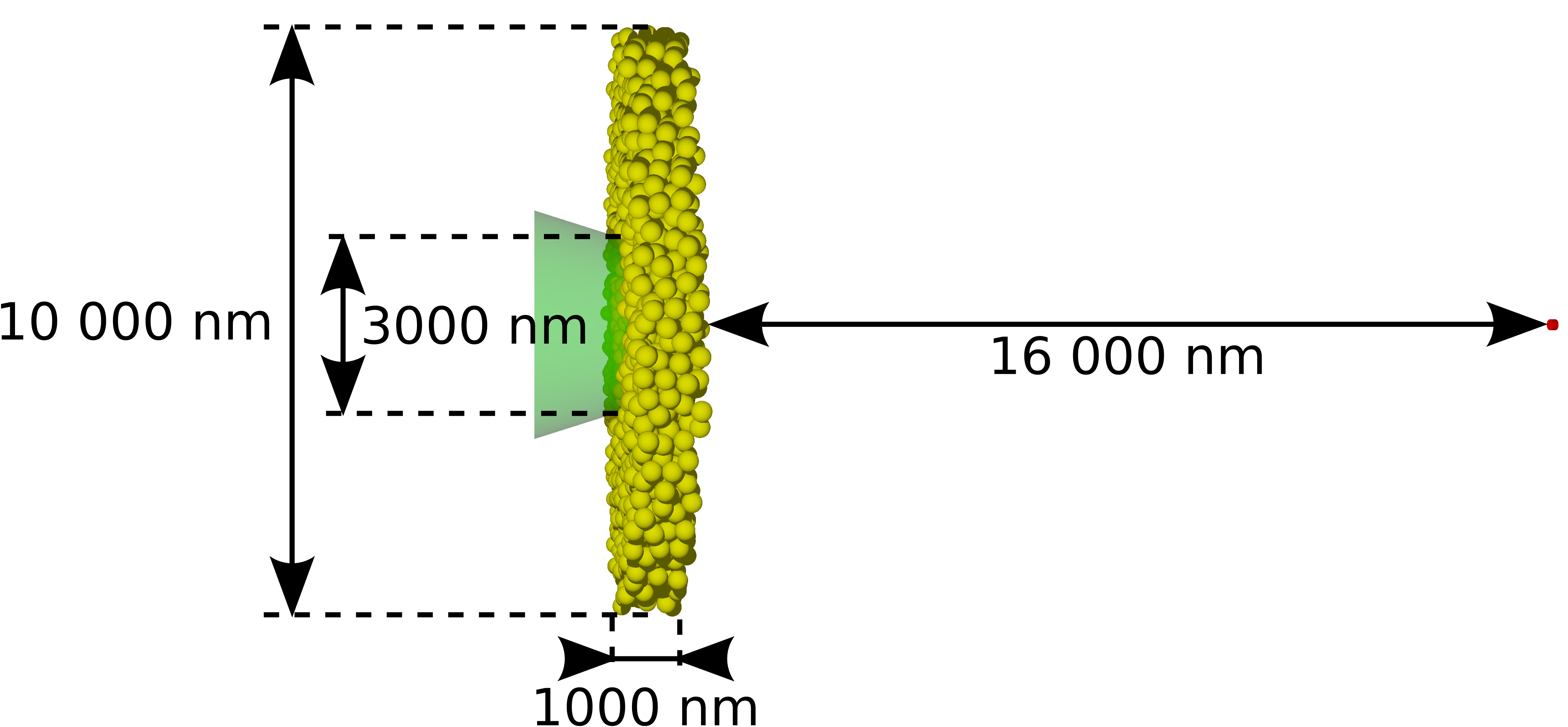}
\caption{MSTM geometry: a Gaussian beam is focused on one side of the sample, and the intensity is calculated in the far field on the red spot (see text for details). The figure is at scale.}
\label{fig:MSTM}
\end{center}
\end{figure}

The transmission calculations of Fig.~\ref{fig:PS-overlstar-simulations}(b) were obtained by using the Multiple Sphere T Matrix code~\cite{Mackowski2011}.
The geometry is shown in Fig.~\ref{fig:MSTM}: we created cylindrical slabs having a diameter of 10000\,nm, a thickness of $L=1000$\,nm containing about 2270 particles with a radius $r=160$\,nm (filling fraction of 50\%).
The five samples were created by using the force-biased algorithm code of Vasili Baranau~\cite{Baranau2014}.

\begin{figure}
\begin{center}
\includegraphics[width=\linewidth]{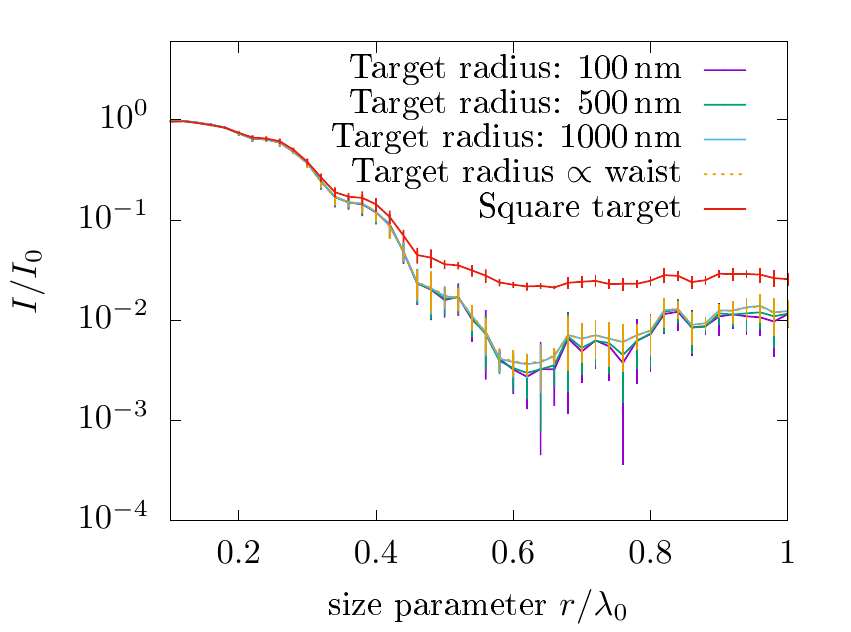}
\caption{Average over five different slabs of the MSTM calculated transmission as a function of the size parameter: influence of the target size on the transmission values. The target is either circular (violet, green, blue and yellow lower curves) or square (red upper curve, square area $10640^2$\,nm$^2$).
Away from the focus, the width of a Gaussian beam depends on the wavelength:
the yellow dashed curve was calculated by taking the radius of the target such that $I_0(r)>I_0(0)/2$.
All error bars correspond to the standard deviation.}
\label{fig:MSTM-trans}
\end{center}
\end{figure}

A Gaussian beam (wavelength $\lambda_0$ between 160 and 1600\,nm)
was focused on one side of the slab (waist $w_0=1500$\,nm), and
$I=\left<E^2\right>$ ---the integral of the intensity on a circle of
radius\footnote{The influence of the target size is shown in
  Fig.~\ref{fig:MSTM-trans}.} 100\,nm placed at a distance of
16000\,nm on the other side of the sample (ten times the largest
wavelength to get rid of any evanescent wave present close to the spheres)--- was calculated\footnote{Five different glass configurations were calculated on the Scientific Compute Cluster (Universität Konstanz) using between 10 and 20 processors for computing time ranging from a few hours to a few days depending on the wavelength.}.
We did the same calculation without any scatterers
($I_0=\left<{E_0}^2\right>$) and the corresponding transmission values
are shown in Fig.~\ref{fig:MSTM-trans} for different target sizes.
These samples are optically thin ($\lambda_0/L$ ranges from 0.16 to 1.6).
We therefore expect most of the photons to be scattered only once.
In this regime, the coherent part of the beam $I_\mathrm{c}$ (i.e. the part of the wave which is not scattered) is attenuated exponentially
\begin{align}
I_\mathrm{c}=I_0\cdot\exp{\left(-\frac{L}{\ell_\mathrm{s}}\right)},
\label{eq:scattering}
\end{align}
where $I_0$ is the incident intensity and $\ell_\mathrm{s}$ is the \emph{scattering} mean free path.

By integrating the intensity on a small surface at such a large
distance on such a thin sample, the major part of it corresponds to the coherent intensity, therefore $I\simeq I_\mathrm{c}$.
This allows us to calculate $\ell_\mathrm{s}$ using Eq.~(\ref{eq:scattering}).

\section{Data analysis with divergence}\label{sec:divergence}
\begin{figure*}
\centering
\begin{tabular}{cc}
\includegraphics[width=0.4\textwidth]{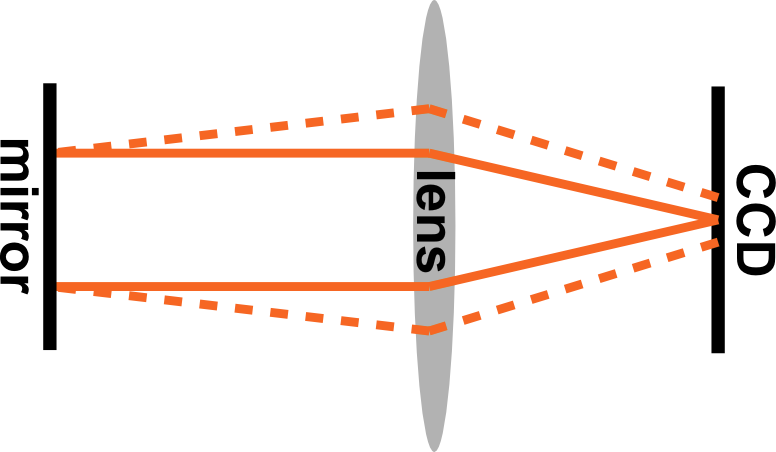}&
\includegraphics[width=0.5\textwidth]{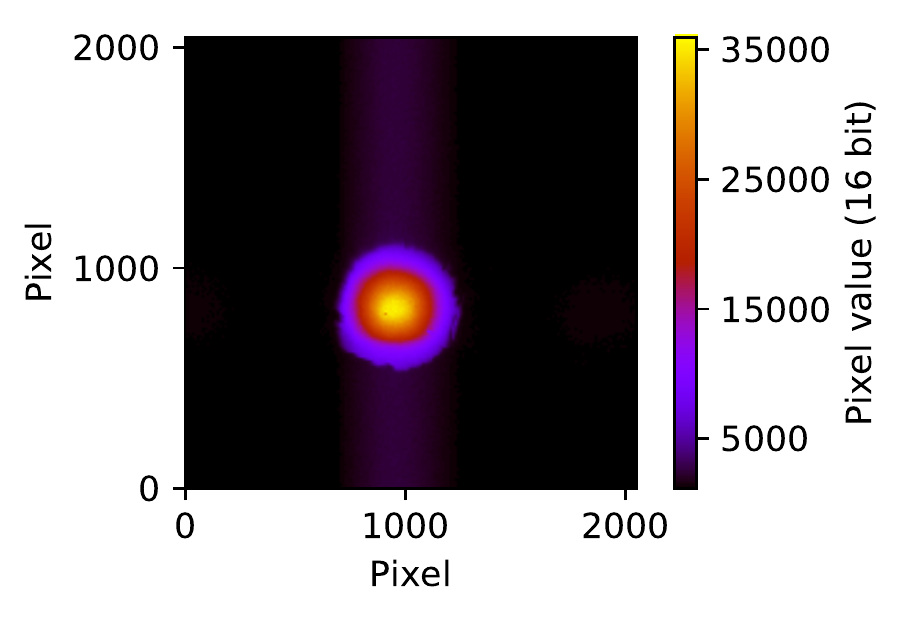}
\end{tabular}
\caption{Sketch of the effect of the divergence on a focused beam and measured image of the divergent spot.}
\label{fig:Spot}
\end{figure*}
Our light source (LED + monochromator) has a divergent beam compared to a coherent laser beam.
This needs to be accounted for in the analysis of the coherent backscattering data.
The backscattered light from the sample diverges slightly before being captured by the lens and imaged in the Fourier space.
This leads to a smeared out (in angles) cone measurement.
The divergence of the parallelized beam from the LED compared to a perfectly parallel beam can be measured by replacing the sample by a mirror.
For a parallel beam this leads to a perfectly focused spot on the CCD if the setup is well adjusted.
The spot imaged from a divergent beam is wider and reflects the angular divergence of the light source (see Fig.~\ref{fig:Spot}).
The spot has a radius of $r \approx 1.4~\mathrm{mm}$.
For the analysis a binary image was generated with a circular spot of the same size centered around the center of the measured CBC. This calibration image was then convoluted with the CBC formula~\cite{MesoPhys} as follows: a two-dimensional image was generated from the CBC formula and transformed with a discrete cosine transform (DCT, a similar function as a discrete Fourier transform but having different boundary conditions). The calibration image was also transformed by a DCT and both functions are multiplied in Fourier space. The result is then transformed back by an inverse DCT. This function is finally used to fit the measured image of the CBC.

%\bibliography{Bibliographie}
%%%%%%%%%%%%%%%% main.bbl
\end{document}